\documentclass[12pt]{article}
\usepackage[utf8]{inputenc}
\usepackage[a4paper, margin={.75in}]{geometry}
\usepackage{amsmath}
\usepackage{amssymb}
\usepackage{multirow}
\usepackage{listings}
\usepackage{graphicx}
\usepackage{pgfplots}
\usepackage{pgf}
\usepackage{mathtools}
\usepackage{tikz}
\usepackage{xcolor}
\usepackage{wrapfig}
\usepackage{bm}
\usepackage{subcaption}
\usepackage[utf8]{inputenc}
\usepackage{soulutf8}
\usepackage{mathabx}
\usepackage{float}
\usepackage{ifthen}
\usepackage{MnSymbol}
\usepackage[hidelinks]{hyperref}
\usepackage{authblk}   
\usepackage{tikz}

\DeclareMathOperator*{\expit}{expit}

\pgfplotsset{compat=1.7}
 \tikzset{bayes/pdf/.style={blue!50!white}}
 \usetikzlibrary{shapes,decorations,arrows,calc,arrows.meta,fit,positioning}

\usetikzlibrary{automata}

\newcommand{\bl}[1]{\textcolor{black}{#1}}

\usepgfplotslibrary{fillbetween}

\usetikzlibrary{plotmarks}

\usetikzlibrary{shapes.swigs}

\tikzset{bayes/pdf/.style={blue!50!white}}

\title{Considerations for Estimating Causal Effects of Informatively Timed Treatments\footnote{\noindent In press at \textit{Epidemiology}.} \footnote{Code available at:\url{https://github.com/stablemarkets/informative_timing} .}  
}

\author[]{Arman Oganisian}
\affil[]{Department of Biostatistics,  Brown University}

 \date{}
\begin{document}
 \maketitle
 \vspace{-.5in}

 \begin{abstract}
Epidemiological studies are often concerned with estimating causal effects of a sequence of treatment decisions on survival outcomes. In many settings, treatment decisions do not occur at fixed, pre-specified followup times. Rather, timing varies across subjects in ways that may be informative of subsequent treatment decisions and potential outcomes. Awareness of the issue and its potential solutions is lacking in the literature, which motivate this work. Here, we formalize the issue of informative timing, problems associated with ignoring it, and show how g-methods can be used to analyze sequential treatments that are informatively timed. As we describe, in such settings, the waiting times between successive treatment decisions may be properly viewed as a time-varying confounders. Using synthetic examples, we illustrate how g-methods that do not adjust for these waiting times may be biased and how adjustment can be done in scenarios where patients may die or be censored in between treatments. We draw connections between adjustment and identification with discrete-time versus continuous-time models. Finally, we provide implementation guidance and examples using publicly available software. Our concluding message is that 1) considering timing is important for valid inference and 2) correcting for informative timing can be done with g-methods that adjust for waiting times between treatments as time-varying confounders.
 \end{abstract}

\newpage

\section{Introduction}

Epidemiological studies often target the causal effect of a sequence of treatment decisions on an event-time outcome. This is especially common in studies of chronic diseases, which are managed with a series of treatment decisions over time rather than administering a single point-treatment. These decisions are not made at fixed, pre-specified followup times. Rather, the timing varies across subjects in ways that are informative of subsequent treatment decisions and outcomes. G-methods provide valid causal inference for time-varying treatments in the presence of time-varying confounding \cite{daniel2013,naimi2016, Robins1986,Robins2000}. However, it is not obvious how these methods can be used in settings with informative, irregularly timed treatments. The purpose of this manuscript is to 1) illustrate biases due to ignoring timing and 2) describe how g-methods can be implemented to adjust for it.

As our central motivating example, will consider recent studies assessing the effects of chemotherapy treatments on survival among patients with pediatric acute myeloid leukemia (AML) \cite{oganisian2024,getz2019a,getz2019b}. These are secondary analyses of the AAML1031 and AAML1053 trials in which enrolled patients moved through multiple treatment courses. At the start of each course, covariates were measured and used to inform the decision to administer anthracyclene chemotherapy (ACT) or some other non-anthracycline chemotherapy (nACT) at that course. The scientific quantity of interest was the proportion of patients that would have survived at least 3 years under various ACT administration strategies. Note ACT inclusion was not randomized in the trial. Due to cardiotoxicity of ACT, ejection fraction (EF) was monitored at each course. Patients with high EF were more likely to be given ACT. An additional complication is that patients did not and, indeed, \textit{could not} initiate each course at pre-set, protocol-defined times. This is because chemotherapy is physically taxing and a subsequent course could only be initiated when a subject was deemed sufficiently stable. According to the AAML1031 trial protocol \cite{aplenc}, ``Patients should progress to the next course of therapy as soon as clinically acceptable...Patients should have time for full or partial recovery from any toxicities experienced in the prior course(s).'' Subjects recovering faster between courses may be systemically better off and tolerate their initial treatment well. As a result, they may both have better future potential survival and be less likely to change their previous treatment. In this sense, the waiting times between treatments can be considered as a kind of time-varying confounder - which we call ``informative timing''.

As another example, consider the use of bisphosphonates in patients diagnosed with osteoporosis \cite{Hayes2021a,Hayes2021b}. Bisphosphonates are a class of drugs used to treat osteoporosis. Since long-term use is thought to lead to increased adverse events, it is common to switch patients to a second-line therapy such as teriparatide or denosumab \cite{Lyu2019}. We can formalize this as a two-course treatment - a first-line treatment decision followed by a second-line treatment decision. The target of inference is cumulative risk of hip fracture or atypical femoral fracture. As in the AML example, different patients switch after different durations of  bisphosphonate therapy. The length of time on first-line therapy may be informative of which drug they switch to as well as outcomes. 

Using AAML1031 as a motivator, we begin by formalizing the problem of informative timing as a type of time-varying confounding. Then, we describe continuous-time approaches for adjusting for informative timing via inverse-probability of treatment weighting (IPTW). Continuous-time causal approaches have been used for structural nested modeling \cite{Lok2008}, structural nested failure time modeling \cite{lok2004}, and marginal structural modeling \cite{Kjetil2011,Ryalen2018}. In epidemiological research, discrete-time causal models are also common and some work has investigated the impact of discretizing continuous data \cite{Zhang2011,Guerra2020,sun2023}. We build on this literature in Section 5 by providing an equivalent discrete-time framing and adjustment for the issue of informative timing.

\section{Potential Outcomes and Informative Timing} \label{sc:standard}
Consider a case with $K$ treatment courses indexed by $k=1,2,\dots, K$. For simplicity of exposition, we consider a case with just $K=2$ courses each of which have two possible treatment options which we denote as $A_k\in\{0,1\}$, respectively. For example, patients undergo $K=2$ courses of chemotherapy and at each course there are two possible chemotherapy agents, ACT ($A_k=1$) or nACT ($A_k=0$). Following the template of the AAML1031 trial, start of followup (time-zero) is the time of the first ($k=1$) treatment course for each patient. Let $L_k$ denote the set of confounders measured at course $k$ just before the treatment decision $A_k$. For concreteness, this could include EF in the AAML1031 context. At course $k=1$, $L_1$ is measured and used to inform ACT assignment, $A_1$. After this course, patients take some time to recover and initiate their second course $W_1$ time units later. If they reach course $k=2$, confounders $L_2$ are once again measured and, along with available information $(A_1, L_1, W_1)$, is used to inform ACT decision $A_2$. After the initiation of the second treatment course, there is some waiting time $W_2$ until death, measured in units of time since start of course $k=2$. Of interest is a survival time outcome, $T>0$. For a subject who survived through two treatment courses, the survival time is the sum of the waiting times $T=W_1 + W_2$. This process is illustrated in the the directed acyclic graph (DAG) in Figure \ref{fig:dagB}. The backdoor path $A_2 \leftarrow W_1 \rightarrow W_2$ is what we refer to as ``informative timing.'' That is, two patients who received $A_2=1$ vs. $A_2=0$ in their second treatment course at different times $W_1$ are, in general, not exchangeable with respect to subsequent survival time $W_2$ - even if the match exactly on $A_1$, $L_1,$ and $L_2$. This may be because subjects who move through their sequence faster (shorter $W_1$) are seen to, say, tolerate their previous treatment especially well and so are more likely to stay on their initial treatment.

In practice, deaths may occur before completion of the sequence so that $W_1$, is the first of either 1) the waiting time until death, $W_{T1}$, or 2) waiting time until the second treatment, $W_{A1}$. We denote this as $W_1 = \min(W_{T1}, W_{A1})$. The corresponding event type indicator is $\delta_1 = I( W_{A1} < W_{T1} )$, where $\delta_1=0$ indicates death following the first treatment course. Until Section \ref{sc:censoring}, we assume there is no censoring. The observed data is given by $D=\{l_{1i}, a_{1i}, w_{1i}, \delta_{1i}\}_{i=1}^n \cup \{l_{2i}, a_{2i}, w_{2i}\}_{i|\delta_{i1}=1}$. As is standard in this literature, we use the overbar notation to denote history (e.g. $\bar L_3=(L_1,L_2,L_3)$) and underbar notation to denote future (e.g. $\underline L_3=(L_3, L_4, \dots, L_K)$).

\begin{figure}
     \begin{subfigure}[b]{0.48\textwidth}
          \centering
           \resizebox{.95\linewidth}{!}{\begin{tikzpicture}
    \node[state] (a1) at (0,0) {$A_1$};
    \node[state] (a2) at (2.5,0) {$A_2$};
    \node[state] (l1) at (-1,2) {$L_1$};
    \node[state] (l2) at (2,2) {$L_2$};
    \node[state] (w1) at (1,1) {$W_1$};
    \node[state] (w2)  at (4,1) {$W_2$};

    \path (l1) edge[ ->] (a1);
    \path (l1) edge[ ->] (w1);
    \path (l1) edge[ ->] (l2);
    \path (l1) edge[bend right=10, ->] (a2);
    \path (l1) edge[bend left=60, ->] (w2);

    \path (a1) edge[ ->] (w1);
    \path (a1) edge[bend right=30, ->] (l2);
    \path (a1) edge[ ->] (a2);
    \path (a1) edge[bend right=60, ->] (w2);

    \path (w1) edge[ ->] (a2);
    \path (w1) edge[ ->] (l2);
    \path (w1) edge[ ->] (w2);

    \path (l2) edge[ ->] (a2);
    \path (l2) edge[ ->] (w2);

    \path (a2) edge[ ->] (w2);
\end{tikzpicture}}  
          \caption{}
          \label{fig:dagB}
     \end{subfigure}
     \begin{subfigure}[b]{0.48\textwidth}
          \centering
          \resizebox{.98\linewidth}{!}{\begin{tikzpicture}

\draw (0,-6) -- (10,-6);
\draw (5,-6) node[below, yshift=-1em]{time};
\draw (0,-6) node[below, yshift=0em, xshift=-.7em]{$j=$};

\draw [dashed] (10,-6) -- (10,.5);
\draw [dashed] (9,-6) -- (9,.5);
\draw [dashed] (8,-6) -- (8,.5);
\draw [dashed] (7,-6) -- (7,.5);
\draw [dashed] (6,-6) -- (6,.5);
\draw [dashed] (5,-6) -- (5,.5);
\draw [dashed] (4,-6) -- (4,.5);
\draw [dashed] (3,-6) -- (3,.5);
\draw [dashed] (2,-6) -- (2,.5);
\draw [dashed] (1,-6) -- (1,.5);
\draw [dashed] (0,-6) -- (0,.5);

\draw (.5,-6) node[below, yshift=-.1em]{$1$};
\draw (1.5,-6) node[below, yshift=-.1em]{$2$};
\draw (2.5,-6) node[below, yshift=-.1em]{$3$};
\draw (3.5,-6) node[below, yshift=-.1em]{$4$};
\draw (4.5,-6) node[below, yshift=-.1em]{$5$};
\draw (5.5,-6) node[below, yshift=-.1em]{$6$};
\draw (6.5,-6) node[below, yshift=-.1em]{$7$};
\draw (7.5,-6) node[below, yshift=-.1em]{$8$};
\draw (8.5,-6) node[below, yshift=-.1em]{$9$};
\draw (9.5,-6) node[below, yshift=-.1em]{$10$};

\draw (-1.5,.5) node[below, yshift=-.6em]{Subject 1};
\draw (0,0pt) -- (7.2,0pt);

\draw (0,3pt) -- (0,-3pt) node[below, xshift = 10pt]{};


\draw (3.7,3pt) -- (3.7,-3pt) node[below]{$W_1$};

\draw[-{Rays[scale=2]} ] (7.2,3pt) node[below, yshift=-.5em]{$T$};

\draw (-1,3pt) node[below, yshift=-1.75em, xshift=-1em]{$V_j=$};
\draw (1,3pt) node[below, yshift=-1.75em, xshift=-1em]{$1$};
\draw (2,3pt) node[below, yshift=-1.75em, xshift=-1em]{$0$};
\draw (3,3pt) node[below, yshift=-1.75em, xshift=-1em]{$0$};
\draw (4,3pt) node[below, yshift=-1.75em, xshift=-1em]{$1$};
\draw (5,3pt) node[below, yshift=-1.75em, xshift=-1em]{$0$};
\draw (6,3pt) node[below, yshift=-1.75em, xshift=-1em]{$0$};
\draw (7,3pt) node[below, yshift=-1.75em, xshift=-1em]{$0$};
\draw (8,3pt) node[below, yshift=-1.75em, xshift=-1em]{$0$};

\draw (-1,3pt) node[below, yshift=-3em, xshift=-1em]{$D_j=$};
\draw (1,3pt) node[below, yshift=-3em, xshift=-1em]{$A_1$};
\draw (2,3pt) node[below, yshift=-3em, xshift=-1em]{$\emptyset$};
\draw (3,3pt) node[below, yshift=-3em, xshift=-1em]{$\emptyset$};
\draw (4,3pt) node[below, yshift=-3em, xshift=-1em]{$A_2$};
\draw (5,3pt) node[below, yshift=-3em, xshift=-1em]{$\emptyset$};
\draw (6,3pt) node[below, yshift=-3em, xshift=-1em]{$\emptyset$};
\draw (7,3pt) node[below, yshift=-3em, xshift=-1em]{$\emptyset$};
\draw (8,3pt) node[below, yshift=-3em, xshift=-1em]{$\emptyset$};

\draw (-1,3pt) node[below, yshift=-4.25em, xshift=-1em]{$X_j=$};
\draw (1,3pt) node[below, yshift=-4.25em, xshift=-1em]{$L_1$};
\draw (2,3pt) node[below, yshift=-4.25em, xshift=-1em]{$\emptyset$};
\draw (3,3pt) node[below, yshift=-4.25em, xshift=-1em]{$\emptyset$};
\draw (4,3pt) node[below, yshift=-4.25em, xshift=-1em]{$L_2$};
\draw (5,3pt) node[below, yshift=-4.25em, xshift=-1em]{$\emptyset$};
\draw (6,3pt) node[below, yshift=-4.25em, xshift=-1em]{$\emptyset$};
\draw (7,3pt) node[below, yshift=-4.25em, xshift=-1em]{$\emptyset$};
\draw (8,3pt) node[below, yshift=-4.25em, xshift=-1em]{$\emptyset$};

\draw (-1,3pt) node[below, yshift=-5.5em, xshift=-1em]{$Y_j=$};
\draw (1,3pt) node[below, yshift=-5.5em, xshift=-1em]{$0$};
\draw (2,3pt) node[below, yshift=-5.5em, xshift=-1em]{$0$};
\draw (3,3pt) node[below, yshift=-5.5em, xshift=-1em]{$0$};
\draw (4,3pt) node[below, yshift=-5.5em, xshift=-1em]{$0$};
\draw (5,3pt) node[below, yshift=-5.5em, xshift=-1em]{$0$};
\draw (6,3pt) node[below, yshift=-5.5em, xshift=-1em]{$0$};
\draw (7,3pt) node[below, yshift=-5.5em, xshift=-1em]{$0$};
\draw (8,3pt) node[below, yshift=-5.5em, xshift=-1em]{$1$};

\draw (-1.5,-3.1) node{Subject 2};
\draw (0,-3) -- (4.4,-3);

\draw (2.3,-2.9) -- (2.3,-3.1) node[below, yshift=-1pt]{$W_1$};
\draw[-{Circle[open, scale=2]}] (4.5,-2.9) node[below, yshift=-5pt]{$C$};

\draw (-1,-7em) node[below, yshift=-1.75em, xshift=-1em]{$V_j=$};
\draw (1,-7em) node[below, yshift=-1.75em, xshift=-1em]{$1$};
\draw (2,-7em) node[below, yshift=-1.75em, xshift=-1em]{$0$};
\draw (3,-7em) node[below, yshift=-1.75em, xshift=-1em]{$1$};
\draw (4,-7em) node[below, yshift=-1.75em, xshift=-1em]{$0$};
\draw (5,-7em) node[below, yshift=-1.75em, xshift=-1em]{$0$};

\draw (-1,-8.2em) node[below, yshift=-1.75em, xshift=-1em]{$D_j=$};
\draw (1,-8.2em) node[below, yshift=-1.75em, xshift=-1em]{$A_1$};
\draw (2,-8.2em) node[below, yshift=-1.75em, xshift=-1em]{$\emptyset$};
\draw (3,-8.2em) node[below, yshift=-1.75em, xshift=-1em]{$A_2$};
\draw (4,-8.2em) node[below, yshift=-1.75em, xshift=-1em]{$\emptyset$};
\draw (5,-8.2em) node[below, yshift=-1.75em, xshift=-1em]{$\emptyset$};

\draw (-1,-9.4em) node[below, yshift=-1.75em, xshift=-1em]{$X_j=$};
\draw (1,-9.4em) node[below, yshift=-1.75em, xshift=-1em]{$L_1$};
\draw (2,-9.4em) node[below, yshift=-1.75em, xshift=-1em]{$\emptyset$};
\draw (3,-9.4em) node[below, yshift=-1.75em, xshift=-1em]{$L_2$};
\draw (4,-9.4em) node[below, yshift=-1.75em, xshift=-1em]{$\emptyset$};
\draw (5,-9.4em) node[below, yshift=-1.75em, xshift=-1em]{$\emptyset$};

\draw (-1,-10.6em) node[below, yshift=-1.75em, xshift=-1em]{$Y_j=$};
\draw (1,-10.6em) node[below, yshift=-1.75em, xshift=-1em]{$0$};
\draw (2,-10.6em) node[below, yshift=-1.75em, xshift=-1em]{$0$};
\draw (3,-10.6em) node[below, yshift=-1.75em, xshift=-1em]{$0$};
\draw (4,-10.6em) node[below, yshift=-1.75em, xshift=-1em]{$0$};
\draw (5,-10.6em) node[below, yshift=-1.75em, xshift=-1em]{$0$};

\end{tikzpicture}}  
          \caption{}
          \label{fig:diag}
     \end{subfigure}
     \caption{(a) DAG depicting scenario discussed in Section \ref{sc:standard}. Here, the time between events explicitly appear on the DAG. $W_1$ is the time from treatment 1 to treatment 2 while $W_2$ is the time from treatment 2 to death, so that $T=W_1+W_2$. It is clear that the time since treatment 1, $W_1$, confounds the effect of $A_2$ on $W_2$. Note that here, unlike in standard settings, the subscripts on the random variables index the treatment course, not time. (b) Figure showing relationship between continuous-time analysis and discrete-time analysis discussed in Section \ref{sc:discrete}. The follow-up time window $[0,\tau]$ is partitioned into $J=10$ equal-width intervals indexed by $j$. $V_j=1$ indicates that a subject had a treatment course at interval $j$. All subjects have their first course at time zero ($j=1$) and a second-course at some random, subject-specific time $W_1$. For example, Subject 1 dies $W_2=T-W_1$ time units after the second course. Covariates ($X_j$) are measured and treatment decision ($D_j$) during these courses - i.e. whenever $V_j=1$. These values are not assessed at other intervals, indicated by $\emptyset$. $Y_j$ is a discretization of the continuous survival time, $T$. It's a monotone indicator that is zero up to the interval of death and one on the interval of death.   }
 \end{figure}

To formalize the causal structure, we define the potential outcome. Let $W_1^{a_1} = \min( W_{A1}^{a_1}, W_{T1}^{a_1} )>0$ be the potential waiting time until the first event after the initiation of the first treatment course had a subject received first-course chemo $a_1$. Similarly, let $\delta_1^{a_1} = I( W_{A1}^{a_1} < W_{T1}^{a_1} )$ be the potential event type - i.e. whether a patient would have died or not before their second course. Thus, the potential survival time for a subject who potentially would have underwent two courses is the sum of the waiting times $T^{a_1, a_2} = W_1^{a_1} + W_2^{a_1, a_2}$. For a subject who would have potentially died before reaching the second treatment $T^{a_1,a_2} = W_1^{a_1}$. More compactly, $T^{a_1,a_2} = W_1^{a_1} + \delta_1^{a_1} W_2^{a_1,a_2}$.

It is helpful to note that the treatment sequence $(a_1, a_2)$ can be viewed as being generated by an underlying dynamic treatment rule (DTR), $r$, that maps the available history at course $k$, $H_k = (\bar A_{k-1}, \bar W_k, \bar L_k, \bar \delta_k )$, to either ACT or nACT. For example, $(a_1=1, a_2=1)$ may be generated by always treating with ACT in the first course $a_1 = r(H_1) = 1$ for any $H_1$. Then, treating with $a_2= r(H_2)= 1$ only if $\delta_2 = 1$ - i.e. if they survive long enough to initiate the second treatment course - and do nothing otherwise. So, contrasts of treatment strategies $(a_1, a_2)$ can be seen as emulations of a target trial where we randomly assign patients at baseline to follow different rules, $r$, for as long as they remain alive. The timing at which this rule is executed to get $a_2 = r(H_2)$ occurs as it naturally would under previous treatment $a_1=r(H_1)$. This mirrors the actual AAML1031 trial protocol which states that patients ``progress to the next course of therapy as soon as clinically acceptable.'' Throughout, the target estimand is $P(T^{1,1} > \tau)$ - the proportion of the target population that would have survived at least $\tau$ time-units had they followed treatment sequence $\bar a_2=(1,1)$.

\section{Adjustment for Informative Timing} \label{sc:adjustment}
Since $P(T^{1,1} > \tau)$ involves a distribution of \textit{potential} survival time, causal identification assumptions are required to map this to the distribution of the \textit{observed} data. With informative timing, one such required assumption is sequentially exchangeable first and second course treatment decisions
\begin{equation} \label{eq:id2}
    \begin{split}
        A_1 & \perp \underline{W}_1^{a_1}, \delta_1^{a_1} \mid L_1=l_1, \\ 
        A_2 & \perp W_2^{a_1,a_2} \mid W_1=w_1, A_1=a_1, L_1=l_1, L_2=l_2, \delta_1=1 .
    \end{split}
\end{equation}
The first statement above asserts that once we condition on $L_1$, it should not be the case that, for example, treated patients are more likely to die before subsequent treatment than untreated patients. The second statement says that among those who have reached the second treatment ($\delta_1=1$), conditioning on timing ($W_1$), previous treatment ($A_1$), and confounder history $(L_1, L_2)$, is sufficient to render the second treatment assignment independent of potential waiting time until death. Positivity assumptions are also required. Specifically, we require that $P(A_1=1 \mid l_1)>0$ for all $l_1$ in the support support and that $P(A_2 = 1 \mid w_{1}, a_1, \bar l_2, \delta_1=1)>0$ for all histories $(w_{1}, a_1, \bar l_2)$ in the support. Intuitively, this second positivity assumption asserts that for each possible value of the history vector $(w_{1}, a_1, \bar l_2)$ among those at-risk for second treatment ($\delta_1=1$), there must be some patients treated with $A_2=1$ and some with $A_2=0$. Otherwise, we would not be able to learn causal effects within that level of the history ($w_1, a_1, \bar l_2$). Since $w_1$ continuous and $\bar l_2$ may include continuous covariates (as in most application), modeling/smoothing is unavoidable since it is rare a treated and untreated unit will have the same exact history ($w_1, a_1, \bar l_2$). Given these assumptions, a g-formula for the quantity of interest is given by
\begin{equation*}
    \begin{split}
        P(T^{a_1, a_2} > & \tau)  = \int_{\tau}^\infty\Big( \int f_{10}(t \mid l_1, a_1) h_{1}(l_1) dl_1 \\
        & + \int \Big[ \int_{0}^t \Big\{ \int f_{W_2}(t-w_1 \mid  w_1, \bar a_2, \bar l_2,  \delta_1 = 1) h_{2}( l_2 \mid a_1, l_1, w_1 )dl_2\Big\}f_{11}(w_1\mid l_1, a_1 ) dw_1 \Big] h_{1}(l_1)  dl_1 \Big) dt  \\
    \end{split}
\end{equation*}
This is a special case of Equation 8.99 in \cite{tsiatis2020} - a derivation is provided in the appendix. Note it is a sum of two terms. The second captures survival from patients who survive through both treatment courses while the first term captures survival for patients who never reach their second course. $h_1$ and $h_2$ denote the density functions of the time-varying covariate while $f_{W_2}$ is the density of the waiting time from course 1 to death. Unlike in typical settings with regularly spaced, protocol-specified treatment times, the g-formula above contains the sub-density function of the waiting time until each event of type $q\in\{0,1\}$, 
$$ f_{1q}(w_1 \mid l_1, a_1) = \lim_{dw \rightarrow 0 } \frac{ P( w_1 \leq W_1 < w_1 + dw, \delta_1=q \mid L_1=l_1, A_1=a_1) }{dw}.$$
The appearance of sub-density functions, commonly encountered in the competing risk literature, is a consequence of death and subsequent treatment ``competing'' to be the next event. They can be modeled directly (e.g. using accelerated failure time models) or indirectly by modeling the corresponding cause-specific hazards \cite{kalbfleisch2011, Austin2016}. An estimate $\hat P(T^{1,1} > \tau)$ can obtained by plugging in model estimates, $\hat h_1, \hat h_2, \hat f_{W_2}, \hat f_{1q}$ into the g-formula and evaluating the integrals via Monte Carlo - an approach known as ``g-computation.'' This outcome modeling approach is especially complex as a competing risk setup is required to capture patients' transition through a sequence of treatments and death in continuous time. In contrast, the Hájek inverse-weighted estimator in \eqref{eq:iptw2} is much simpler as it only requires models for the two treatment decisions:

\begin{equation} \label{eq:iptw2}
\hat P(T^{1,1} > \tau) = \frac{ \sum_{i | \delta_{1i}=0} A_{1i} I(W_{1i} > \tau) \hat \omega_i + \sum_{i | \delta_{1i}=1} A_{1i}A_{2i} I(W_{1i} + W_{2i} > \tau) \hat \omega_i  }{ \sum_{i | \delta_{1i}=0} A_{1i} \hat \omega_i +  \sum_{i | \delta_{1i}=1} A_{1i}A_{2i} \hat \omega_i }.   
\end{equation}

Notice there are two terms in both the numerator and denominator. The first sums over contributions from patients who only received one treatment before death and the second sums over contributions from patients who received two treatments before death. The weights are given by $\hat \omega_i = 1/\hat \pi_i$ where for subjects with $\delta_{1i}=0$, the propensity score estimate is
\begin{equation} \label{eq:pscore_12}
    \hat \pi_i =\hat P( A_1 = 1 \mid L_1=l_{1i}).
\end{equation}
For subjects with $\delta_{1i}=1$ the propensity score estimate is
\begin{equation} \label{eq:pscore_122}
    \hat \pi_i = \hat P(A_2 = 1 \mid W_1=w_{1i}, A_1 = 1, L_2 = l_{2i}, L_1=l_{1i}, \delta_1=1) \hat P( A_1 = 1 \mid L_1=l_{1i}). 
\end{equation}
Here, the conditional probability of the second treatment is estimated using only patients who survived long enough to receive it ($\delta_1=1$), and this is again justified by the second line of the exchangeability condition in \eqref{eq:id2}. This estimator allows for the treatment probability at the second course to depend on time elapsed since the first treatment, $W_1$. The top panel of Table \ref{tab:simres} displays results from a simulation in which patients move through $K=2$ courses and initiate their second course at some random time, $W_1$, that depends on prior treatment and covariates and also informs subsequent treatment, $A_2$, and survival, $W_2$. 
The ``Unadjusted IPTW'' estimator refers to the estimator in \eqref{eq:iptw2} where the propensity score weights do not adjust for $W_1$. Notice that it produces a biased points estimate (8.76\%) and 95\% intervals with under-coverage (80.3\%). On the other hand, adjusting for $W_1$ in \eqref{eq:iptw2} - labeled ``Adjusted IPTW'' - alleviates this bias. The ``Naive'' estimator that subsets to subjects who survived through both courses and were treated in both is obviously severely biased. The bias in the naive estimator shows the importance of properly accommodating deaths between treatments as well as adjustment for informative timing.

\section{Censoring Between Treatment Courses} \label{sc:censoring}

In the case where patients drop out of a study, survival time is censored. Accordingly, the observed waiting time after the first treatment is $W_1=\min(W_{T1}, W_{A1}, C_1)$, where $C_1$ is the waiting time until censoring. Now, there are three possible events of type $\delta_1 \in\{1,0,-1\}$, indicating a subsequent treatment event, death event, or censoring event respectively.  Similarly, for those who received a second treatment, the observed waiting time is the first of either the waiting time from second treatment to death, $W_{T2} = T-W_1$, or the waiting time from second treatment to censoring, $C_2$. We denote this as $W_2 = \min(W_{T2}, C_2)$. The associated indicator is $\delta_2 = I(W_{T2} < C_2)$. The observed data, then, is then $D=\{l_{1i}, a_{1i}, w_{1i}, \delta_{1i}\}_{i=1}^n \cup \{l_{2i}, a_{2i}, w_{2i}, \delta_{2i}\}_{i|\delta_{i1}=1}$. We also define an indicator for either death or treatment (i.e. not censoring) between treatment 1 and 2, $\delta_{1i}^E = I(\delta_{1i} \neq -1) $. The methods of the previous section can be modified under an additional non-informative censoring assumption
\begin{equation}
    \begin{split}
        C_1 & \perp W_1^{a_1}, W_2^{a_1, a_2} \mid A_1, L_1, \\
        C_2 & \perp W_2^{a_1, a_2} \mid A_1, A_2, L_1, L_2, W_1, \delta_1=1.
    \end{split}
\end{equation}
This states that, after each treatment, censored patients should not be any more or less likely to experience the next event any sooner/later than uncensored subjects conditional on the available information. In addition to exchangeability, we also require additional positivity assumptions about the censoring mechanism. Specifically, there must be some positivity probability of remaining uncensored long enough to realize a second event, 
$$ S_{C1}(w_1 \mid A_1, L_1) = P(C_1 > w_1 \mid A_1, L_1) > 0.$$
for each $w_1$ and all levels of $(A_1, L_1)$. Here $S_{C1}(w_1 \mid -)$ is the ``survival'' function of the waiting time to censoring. If this assumption did not hold, then there would be some subgroup of patients who would always be censored before an event - making it impossible to learn about counterfactual times within this subgroup. Similarly, 
$$ S_{C2}(w_2\mid A_1, A_2, L_1, L_2, W_1, \delta_1=1) = P(C_2 > w_2 \mid A_1, A_2, L_1, L_2, W_1, \delta_1=1) > 0.$$
among patients who reached the second treatment, there must be some chance of remaining uncensored long enough to reach a death event. 

\begin{table}[]
\centering
\begin{tabular}{llrrrr}
 Setting & Method & \multicolumn{1}{l}{\%Bias} & \multicolumn{1}{l}{rel. MSE} & \multicolumn{1}{l}{Width} & \multicolumn{1}{l}{Coverage} \\ \hline
\multicolumn{1}{l|}{\multirow{3}{*}{No censoring}} & \multicolumn{1}{l|}{Adjusted IPTW} & 0.294  & 1      & 0.05   & .942 \\
\multicolumn{1}{l|}{} & \multicolumn{1}{l|}{Unadjusted IPTW}                            & 8.76   & 2.51   & 0.05   & .803 \\
\multicolumn{1}{l|}{} & \multicolumn{1}{l|}{Naive}                                      & 45.50  & 38.5   & 0.06   & .064 \\ \hline
\multicolumn{1}{l|}{\multirow{3}{*}{Censoring}} & \multicolumn{1}{l|}{Adjusted IPTW}    & 0.03   & 1      & 0.07   & .947 \\
\multicolumn{1}{l|}{} & \multicolumn{1}{l|}{Unadjusted IPTW}                            & 214.00 & 484.00 & 0.40   & .119 \\
\multicolumn{1}{l|}{} & \multicolumn{1}{l|}{CC-IPTW}                                    & 22.22  & 4.26   & 0.04   & .012 \\ \hline
\end{tabular}
\caption{Bias and MSE of various point estimator methods for $P(T^{1,1}>15)$. Bias is calculated as average difference between estimate and truth value. \%Bias is calculated as Bias divided by the truth value. We also report average width and coverage probability of the 95\% confidence interval. Intervals were computed using bootstrap via 1000 replicates using the percentile method. Results are across 1,000 simulation data sets, each of sample size $n=2000$. We use 500 bootstrap replicates. More details on the simulation are provided in the Appendix.}
\label{tab:simres}
\end{table}

If these conditions hold, then we can estimate $P(T^{1,1}>\tau)$ as
\begin{equation} \label{eq:iptw3}
    \hat P(T^{1,1}>\tau) = \frac{ \sum_{i | \delta_{1i}\neq1} A_{1i} \delta_{1i}^E I( W_{1i} > \tau) \hat \omega_i + \sum_{i | \delta_{1i}=1} A_{1i}A_{2i} \delta_{2i} I(W_{1i} + W_{2i}>\tau) \hat \omega_i  }{ \sum_{i | \delta_{1i}\neq1} A_{1i} \delta_{1i}^E \hat \omega_i +  \sum_{i | \delta_{1i}=1} A_{1i}A_{2i} \delta_{2i} \hat \omega_i }.
\end{equation}

The estimator only includes contributions from patients who follow sequence $(A_1=1, A_2=1)$ for as long as they are at-risk. That is, $A_1 \delta^E = 1$ only for uncensored subjects who underwent just one treatment course before death and $A_1A_2\delta_2=1$ only for for uncensored subjects who underwent two treatment courses. The weights must now correct for the fact that this subset of patients are not representative of the target population both in terms of treatment \textit{and} censoring probability and so they become $\hat \omega_i = 1/( \hat \pi_i\cdot \hat \eta_i)$. Here, $\hat \pi_i$ is the estimated propensity score as given in Equations \eqref{eq:pscore_12} and \eqref{eq:pscore_122} in the previous section. The term $\hat \eta_i$ is the estimated probability of subject $i$'s observed censoring history. For subjects who underwent one treatment before death, this is given by the conditional probability that the waiting time to censoring exceeds the observed waiting time, $w_{1i}$,
$$ \hat  \eta_i = \hat S_{C1}(w_{1i} \mid A_1=1, L_1 = l_{1i}). $$
Similarly, for subjects who underwent two treatment courses, this is given by the conditional probability that the waiting time from treatment 1 to censoring exceeds the observed waiting time until treatment 2, multiplied by the conditional probability of the waiting time from treatment 2 to censoring exceeding the observed waiting time from treatment 2 to death
$$ \hat \eta_i = \hat S_{C2}( w_{2i} \mid A_1=1, A_2=1, L_1=l_{1i}, L_2=l_{2i}, W_1 = w_{1i}, \delta_1=1) \hat S_{C1}( w_{1i} \mid A_1=1, L_1 = l_{1i}). $$
Note that computation of these weights require estimates of the censoring probabilities, $ \hat S_{C1}$ and $ \hat S_{C2}$. The bottom panel of Table \ref{tab:simres} displays results from a simulation in which patients may die or be censored before undergoing a second treatment. Both ``Adjusted IPTW'' and ``Unadjusted IPTW'' methods implement \eqref{eq:iptw3}. The former adjusts for $W_1$ when computing $\hat \pi_i$ and $\hat \eta_i$, whereas the latter does not. Again, failing to adjust for timing results in bias and under-coverage. The method ``CC-IPTW'' method implements the estimator \eqref{eq:iptw2} using only subjects who are uncensored. This essentially fails to weight for censoring and thus leads to biased estimates. The bias of CC-IPTW emphasizes the need to handle censoring by including contributions of both patients who die after completion of the sequence as well as those who die in between treatments. Ignoring any of these complexities may result in poor estimates.

\section{Adjustment with Discrete-Time Models} \label{sc:discrete}

We now describe the above concepts within a discrete-time framework. We show that causal estimands, identification results, and the need to adjust for informative timing are agnostic to the choice of discrete versus continuous time modeling and so the main takeaways are analogous. For simplicity, we consider a case with no censoring until the end.

Again, the target of inference is $P(T^{1,1}>\tau)$. We start by partitioning the time window $[0,\tau]$ into $J$ disjoint, equal-length intervals indexed by $j= 1,2,\dots, J$ with interval $j=1$ being the baseline. Let $D_j\in\{\emptyset, 0, 1\}$, be the treatment decision made in interval $j$ where $\emptyset$, 0, and 1 denote no treatment, decision to treat with nACT, and decision to treat with ACT, respectively. We let $V_j\in\{0,1\}$ be an indicator of whether a treatment course occurred at interval $j$ and we let $X_j$ denote covariates measured in interval $j$. We will let $X_j=\emptyset$ denote that covariates were not assessed in interval $j$. Finally, we let $\{Y_j\}_{j=1}^J$ be a monotone indicator trajectory which discretizes the continuous survival time outcome, $T$. Specifically, $Y_j=0$ for all intervals up to the interval of death and $Y_j=1$ for all intervals on and after the interval of death, as shown in Figure \ref{fig:dagB}

The continuous-time analysis of the two-course treatment sequence of the previous sections correspond to an analysis of these discrete-time variables under the following constraints: First, covariates are assessed and treatments decisions are made only during a treatment course: i.e., $V_j = 0 \implies D_j = \emptyset$ and $X_j=\emptyset$ with probability 1 for all $j$. Second, all subjects initiate the first course at baseline so that $V_1=1$ for all subjects. Third, at a single, random future interval $S \in \{2,\dots, J+1 \}$, $V_{S}=1$. That is, the treatment of interest consists of only two courses (one post-baseline course) and different patients may initiate this second course at different intervals, $S$. The value $S = J+1$ indicates that the patient did not initiate the second course within $\tau$. Thus there is a 1-1 correspondence between the random variable $S$ and the length-$J$ random vector $\bar V_J$. For example, $S=5$ implies $\bar V_J$ contains a 1 in the $1^{st}$ and $5^{th}$ entries, and is zero in all other entries - and vice versa. To link this back to the continuous-time notation explicitly, the first course is initiated at the first interval $j=1$ so $V_j=1$ and we measure covariates $X_1 = L_1$ and make decision $D_1 = A_1\in\{0,1\}$. The second course is initiated at $W_1$ which is in interval $S$ so that $V_{S}=1$. At this interval, covariates are collected $X_{S} = L_2$ and a second treatment decision is made, $D_{S} = A_2 \in\{0,1\}$. This relationship between the discrete and continuous-time variables is illustrated in Figure \ref{fig:diag}.

Now we consider the discrete-time potential outcomes. Just as in Section \ref{sc:standard}, the vector of observed decisions, $\bar d_J$, can be framed as the output of a dynamic treatment rule (DTR), $r_j(V_j, Y_j; \bar a_2)$, where treatment course initiation and survival status are tailoring variables. Specifically, at the first interval, $V_1=1$ and $Y_1=0$ and so $d_1=r_1(1, 0; \bar a_2) = a_1$. For later intervals $j>1$ the rule $r_j$ makes decision $a_2$ if a patient is alive and has their second course at interval $j$. If they are either dead or a treatment course does not occur at interval $j$, then no decision is made. Formally,
\begin{equation}
    r_j(V_j, Y_j;\bar a_2)=
    \begin{cases}
        a_2, & \text{if}\ V_j=1, Y_j=0 \\
        \emptyset, & \text{otherwise}
    \end{cases}
\end{equation}
In the DTR literature, $\bar a_2$ would be called the ``parameter'' of the rule. The sequence of rules $r=(r_1, r_2,\dots, r_J)$ explicitly connects the treatment decisions in continuous-time $(a_1, a_2)$ with the discretized decision sequence $ \bar d_J = (d_1, d_2, \dots, d_J)$. That is, we should really write $\bar d_j^{r}$ as it is the sequence of decisions made by following rule $r$ up through interval $j$, but we avoid explicit indexing by $r$ for compactness. The DTR formulation clearly distinguishes between the aspects of the decision rule we \textit{can} specify in a trial protocol from what we \textit{cannot} specify in a protocol. In trials like AAML1031, while we can specify \textit{which} decisions are made, $\bar a_2$, we cannot specify \textit{when} they are made, $S$. \\

We let $Y_j^{\bar d_j}\in\{0,1\}$ denote the potential death indicator with zero indicating a patient would have remained alive through interval $j$ and one denoting dead in interval $j$ had they followed treatment sequence $\bar d_j$. The terminal nature of death implies that $Y_{j-1}^{\bar{d}_{j-1}} = 1 \implies Y_{j}^{\bar{d}_{j}} = 1$ with probability 1 for all $j$. The target estimand of $P(T^{1,1} > \tau)$ in this discrete-time setting is given by
\begin{equation} \label{eq:dischaz}
    \begin{split}
        P(T^{a_1,a_2} > \tau) & = P(Y_1^{d_1} = 0, Y_2^{\bar{d}_2} = 0, \dots,   Y_J^{\bar{d}_J} = 0) \\
        & = \prod_{j=1}^J \big(   1 - P(Y_j^{\bar d_j} = 1 \mid  Y_{j-1}^{\bar d_{j-1}} = 0 ) \big) \\ 
    \end{split}
\end{equation}
where $P(Y_j^{\bar d_j} = 1 \mid  Y_{j-1}^{\bar d_{j-1}} = 0 )$ is the marginal discrete-time hazard as discussed in \cite{Hernan2000}. Identification of the discrete-time hazard requires a condition analogous to \eqref{eq:id2}
\begin{equation} \label{eq:id3}
    D_j \perp \underline Y_{j+1}^{\bar d_j} \mid \bar X_j = \bar x_j,  \bar D_{j-1} = \bar d_{j-1}, \bar V_j = \bar v_j, \bar Y_{j} = 0 
\end{equation}
This is the usual sequential exchangeability assumption where $X_j$ and $V_j$ are treated as observed time-varying covariates. \bl{Graphically, this asserts that $(\bar X_j ,  \bar D_{j-1}, \bar V_j, \bar Y_{j})$ d-separates $D_j$ and $\underline Y_{j+1}^{\bar d_j}$ on a single-world intervention graph (See Appendix A4)}. The assumption holds trivially except at $j=1$ and $j=S$. This is because $V_j=0$ for all other $j$, and so $D_j=\emptyset$ deterministically. At $j=1$, everyone initiates their first course $V_1=1$ and the covariates consist of $X_1=L_1$ and so \eqref{eq:id3} simplifies to

$$  A_1 \perp \underline Y_{2}^{\bar d_1} \mid L_1 = l_1, V_1 = 1, \bar Y_{1} = 0 $$

That is, among those at-risk at the first interval ($Y_{1} = 1$) who initiate the first treatment course at the first interval ($V_1 = 1$) the treatment decision, $D_1\in\{0,1\}$ is unrelated to future potential survival trajectory. Replacement of $D_1$ with $A_1$ is justified since they are linked by $r$, $D_1 = r_1(1,0)=A_1$. Notice that this is equivalent to the first line of \eqref{eq:id2}.  Similarly, at the interval of the second treatment course, $j=S$, the covariate history $\bar X_{S}$ only consists of two non-null values, $(L_1, L_2)$. And $\bar D_{j-1}$ only consists of one non-null value, $A_1$. Thus, \eqref{eq:id3} simplifies to 
$$  A_{2} \perp \underline Y_{j+1}^{\bar d_{j}} \mid L_1 = l_1, L_{2} = l_{2}, A_1 = a_1, S = j,  Y_{j} = 0 $$

Replacing conditioning on $\bar V_{j}$ in \eqref{eq:id3} with $S$ is justified by the aforementioned 1-1 correspondence between them. This conditional independence statement asserts that among those at risk ($ Y_{j} = 0$) patients with different $A_2$ who have the same elapsed time since previous treatment, $S$, same covariates $L_1$, $L_{2}$, and same prior treatment $A_1$ are exchangeable. This is analogous to the second line of \eqref{eq:id2}. Just as in \eqref{eq:id2} we require conditioning on the \textit{time elapsed} since the first course $W_1$ to adjust for informative timing, here we require conditioning on the discrete-time analogue - the \textit{number of intervals elapsed} since the last course $S$. \\

\begin{table}[h] \centering
\begin{tabular}{lll}
\multicolumn{3}{c}{Data Summary Statistics}                             \\ \hline
                                      & Course $k=1$   & Course $k=2$   \\ \hline
\multicolumn{1}{l|}{}                 & $n=600$        & $n=416$        \\ \hline
\multicolumn{1}{l|}{Treatment, $A_k$} &                &                \\
\multicolumn{1}{r|}{ACT}              & $235 \ (60.8\%)$ & $256 \ (61.5\%)$ \\
\multicolumn{1}{r|}{nACT}             & $365 \ (39.2\%)$ & $160 \ (38.5\%)$ \\
\multicolumn{1}{l|}{EF, $L_k$}        &                &                \\
\multicolumn{1}{r|}{Normal}           & $317 \ (52.8\%)$ & $181 \ (43.5\%)$ \\
\multicolumn{1}{r|}{Low}              & $283 \ (47.2\%)$ & $235 \ (56.4\%)$ \\ \hline
\end{tabular}
\caption{Summary of synthetic data consisting of $600$ patients moving through two treatment courses. At each course, the table lists count and percentage (of at-risk set) of patients receiving ACT ($A_k=1$) vs. nACT ($A_k=0$) and patients who have normal ($L_k=0$) versus low ($L_k=1$) ejection fraction (EF). 416 patients reach the second treatment course. Of the remaining 184 subjects, 129 either die and 55 are censored before initiating the second course. Median waiting time until the second course is 4.2 months and median overall survival time is 7 months.}
\label{tab:sumtab}
\end{table}

Under these identification assumptions (along with analogous sequential consistency and positivity assumptions), the Hájek inverse-probability of treatment weighted estimator for the discrete-time hazard at interval $j$ is
\begin{equation} \label{eq:dt_iptw}
    \hat P(Y_j^{\bar d_j} = 1 \mid  Y_{j-1}^{\bar d_{j-1}} = 0 ) = \frac{ \sum_{i=1}^n I( D_{i,j} = d_{i,j})  Y_{i,j}(1-Y_{i,j-1})  \hat \omega_{i,j} 
  }{ \sum_{i=1}^n I(D_{i,j}= d_{i,j} )  (1-Y_{i,j-1}) \hat \omega_{i,j}  } 
\end{equation}
See Equation 39 in \cite{Young2020} which presents this estimators in settings with competing events. Extensions to the DTR settings are given in \cite{Neugebauer2012} and \cite{tsiatis2020} (Section 8.3). The indicator $I(D_{i,j}= d_{i,j} )$ selects person-period contributions with decisions, $D_{i,j}$, consistent with the DTR $r$. The time-varying weights are $\hat \omega_{i,j} = 1/\hat \pi_{ij}$, where the estimated time-varying propensity score is
$$ \hat \pi_{i,j} = \prod_{u=1}^j \hat P(D_u = d_{i,u} \mid \bar D_{u-1} = \bar d_{i,u-1}, \bar X_u = \bar{X}_{iu}, \bar V_u = \bar V_{iu}, \bar Y_u = 0 ) $$
In our setting, this formula simplifies. Since $D_j=\emptyset$ whenever $V_{j}=0$ with probability 1 and $V_{j}=0$ at only $j=1$ and $j=S$, the product in $\hat \pi_{i,j}$ is comprised only of at most two factors, not $j$ factors. This is because all other factors in the product of $\hat \pi_{i,j}$ are $P(D_j = \emptyset \mid -, V_j=0 )=1$. The weight for person-time intervals $1 \leq j < S $ is 

$$ \hat \pi_{i,j} = \hat P(A_1 = 1 \mid L_{i,1}, V_1=1, Y_1 = 0 ) $$

For subject $i$  who has had both treatment courses by interval $k$, the weight person-time interval $j\geq S$ is 

$$ \hat \pi_{i,j} = \hat P(A_1 = a_1 \mid L_{i,1},  V_1 = 1, Y_1 = 0 ) \hat P(A_{2} = a_2 \mid L_{i,2}, L_{i,1}, A_1 = a_1, S=j, V_1 = 1, Y_{j} = 0 )  $$

Notice that these are the same (up to a discretization) as \eqref{eq:pscore_12} and \eqref{eq:pscore_122}. Just as in \eqref{eq:pscore_122} we control for the \textit{time elapsed} since the first course $W_1$, here we control for the \textit{number of intervals elapsed} since the last course, $S$. While $\hat P(Y_j^{\bar d_j} = 1 \mid  Y_{j-1}^{\bar d_{j-1}} = 0 )$ can be computed directly from the formula above for each $j$, it is commonly computed via a saturated marginal structural cox model (MSM) \cite{Hernan2000}. That is, the discretized data are arranged in long person-interval format and a logistic regression model with interval-specific intercepts is fit for $Y_j$ using the $\hat \omega_{i,k}$ as weights. In the case of censoring, the time-varying weights $\hat \omega_{i,j}$ will be the inverse of the product of both the treatment probabilities and the probability of remaining uncensored. The latter is estimated via another pooled logistic regression on a monotone censoring indicator \cite{Hernan2000}.

\begin{figure}[h!]
     \begin{subfigure}[b]{0.48\textwidth}
          \centering
           \resizebox{.98\linewidth}{!}{\includegraphics[scale=1]{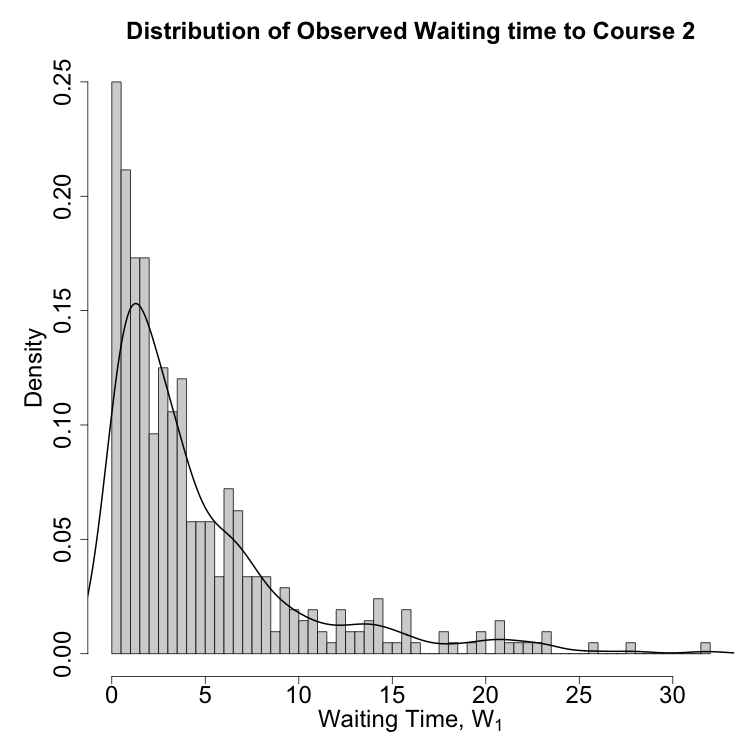}}  
          \caption{}
          \label{fig:wtdist}
     \end{subfigure}
     \begin{subfigure}[b]{0.48\textwidth}
          \centering
          \resizebox{.98\linewidth}{!}{  \includegraphics[scale=1]{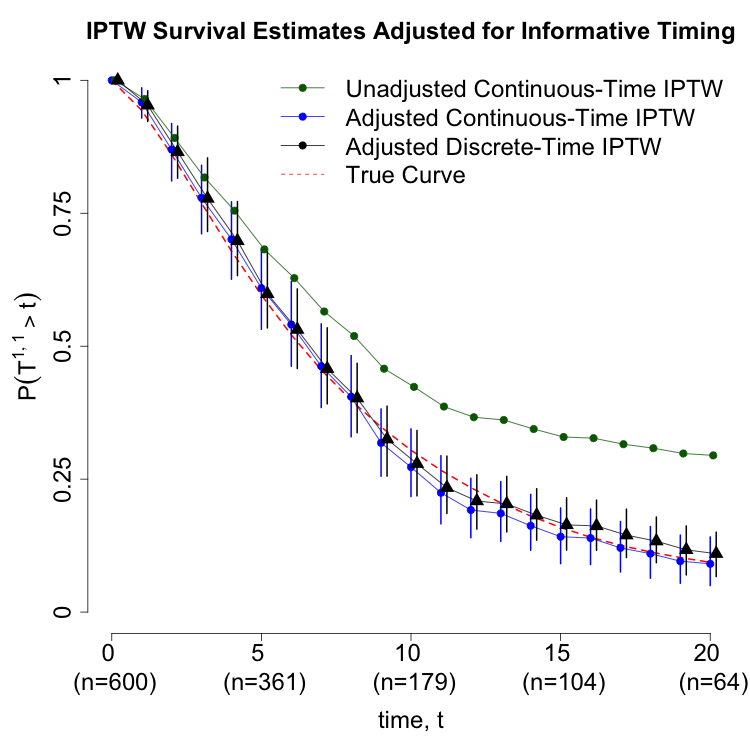}}  
          \caption{}
          \label{fig:survplot}
     \end{subfigure}
     \caption{(a) Distribution of the observed waiting time, $W_1$, to the second treatment course among patients who reached it. Note that there is variation in the timing of the second treatment and a long tail of patients who take especially long to recover from their first course. (b) Estimates of potential survival rate, $P(T^{1,1} > \tau)$, plotted for $\tau\in\{0,1,2,\dots, 20\}$. Estimates are point-wise and are connected by lines just for ease of visualization. The blue and green curves are computed via Equation \eqref{eq:iptw3}. In blue, we adjust for $W_1$ in the propensity score weights and in green we do not. For comparison, the black curve shows the $W_1$-adjusted curve using the discrete-time approach in Equation \eqref{eq:dt_iptw}. }
 \end{figure}

To compare both approaches, we simulated a dataset mimicking AAML1031, summarized in Table \ref{tab:sumtab}. For simplicity of illustration, we have a binary time-varying covariate corresponding to normal ($L_k=0$) versus low ($L_k=1$) ejection fraction (EF). In this dataset, 416 patients reach the second treatment course. The rest die or are censored beforehand. The distribution of the waiting time until this second course is given in Figure \ref{fig:wtdist}. Figure \ref{fig:survplot} depicts estimates of $P(T^{1,1} > \tau)$ at points $\tau\in0,1,2,\dots, 20$. It depicts estimates using the continuous-time formula in \eqref{eq:iptw3} where we include waiting time $W_1$ as a covariate in the second-stage propensity score and censoring models (in blue) and where we exclude it (in green). In black, we show estimates using the discrete-time adjustment in Equation \eqref{eq:dt_iptw}. Both adjustment procedures lead to estimates close to truth and perform similarly. On the other hand, failing to adjust for $W_1$ yields an estimate that is off. More details on software implementation are provided in the Appendix and the code\footnote{\url{https://github.com/stablemarkets/informative_timing}} is made available online.

\section{    \bl{Discussion}     }

\bl{The concepts discussed here relate to the literature on irregular visit/observation processes. Continuous-time \cite{Rosenbeck2004, Cook2019, Lim2016} and discrete-time approaches \cite{Hernan2009} have been developed both within and outside of a formal causal inference setting. In the causal context with discrete-time models, these consider the observation process (roughly – but not exactly – analogous to $\{V_j\}$ in Section 5) to be intervenable and target causal effects of a specified joint treatment-observation strategy $(\bar d_J, \bar v_J)$. Person-intervals are artificially censored when they deviate from the specified strategy. The remaining person-intervals are re-weighted using the \textit{joint} probability of their treatment \textit{and} observation process. However in motivating settings, such as the AAML1031 trial, considered here, the timing of the treatment courses are not intervenable so such contrasts are not clinically relevant. Such weighting schemes may lead to unstable estimates as there may be very few person-intervals consistent with a specified timing strategy $\bar{v}_J$ in such settings. Therefore, we instead work within an alternative framework \cite{tsiatis2020, Hager2018} which views waiting times between successive events as potential outcomes of previous treatment decisions rather than a process controllable by design. It is possible, however, that these procedures could also be equivalently motivated via a stochastic intervention that sequentially sets each component of $\bar{V}_J$ according to its natural, time-varying distribution. A formal connection is an interesting area of future inquiry.}

\bibliographystyle{plain}
\bibliography{refs} 

\begin{thebibliography}{10}

\bibitem{aplenc}
Richard Aplenc and Children's~Oncology Group.
\newblock {AAML1031: A Phase III Randomized Trial for Patients with de novo AML
  using Bortezomib and Sorafenib for Patients with High Allelic Ratio
  FLT3/ITD}.
\newblock {\em clinicaltrials.gov}, 2017.

\bibitem{Austin2016}
Peter~C. Austin, Douglas~S. Lee, and Jason~P. Fine.
\newblock Introduction to the analysis of survival data in the presence of
  competing risks.
\newblock {\em Circulation}, 133(6):601--609, 2016.

\bibitem{Cook2019}
Richard~J Cook and Jerald~F Lawless.
\newblock Independence conditions and the analysis of life history studies with
  intermittent observation.
\newblock {\em Biostatistics}, 22(3):455--481, 11 2019.

\bibitem{daniel2013}
R.M. Daniel, S.N. Cousens, B.L. De~Stavola, M.~G. Kenward, and J.~A.~C. Sterne.
\newblock Methods for dealing with time-dependent confounding.
\newblock {\em Statistics in Medicine}, 32(9):1584--1618, 2013.

\bibitem{Guerra2020}
Steve Ferreira~Guerra, Mireille~E. Schnitzer, Amélie Forget, and Lucie Blais.
\newblock Impact of discretization of the timeline for longitudinal causal
  inference methods.
\newblock {\em Statistics in Medicine}, 39(27):4069--4085, 2020.

\bibitem{getz2019a}
Kelly~D Getz, Lillian Sung, Robert~B. Gerbing, Todd~A. Alonzo, Yimei Li,
  Yuan-Shung~V. Huang, Kasey~J Leger, Jessica~A. Pollard, Todd~M Cooper,
  Edward~Anders Kolb, Alan~S Gamis, and Richard Aplenc.
\newblock Occurrence and resolution of anthracycline cardiotoxicity and impact
  on treatment outcomes among children treated on the aaml1031 clinical trial:
  A report from the children's oncology group.
\newblock {\em Blood}, 134:331, 2019.

\bibitem{getz2019b}
Kelly~D. Getz, Lillian Sung, Bonnie Ky, Robert~B. Gerbing, Kasey~J. Leger,
  Allison~Barz Leahy, Leah Sack, William~G. Woods, Todd Alonzo, Alan Gamis, and
  Richard Aplenc.
\newblock Occurrence of treatment-related cardiotoxicity and its impact on
  outcomes among children treated in the aaml0531 clinical trial: A report from
  the children's oncology group.
\newblock {\em Journal of Clinical Oncology}, 37(1):12--21, 2019.

\bibitem{Hager2018}
Rebecca Hager, Anastasios~A. Tsiatis, and Marie Davidian.
\newblock Optimal two-stage dynamic treatment regimes from a classification
  perspective with censored survival data.
\newblock {\em Biometrics}, 74(4):1180--1192, 05 2018.

\bibitem{Hayes2021b}
K.~N. Hayes, N.~He, K.~A. Brown, A.~M. Cheung, D.~N. Juurlink, and S.~M.
  Cadarette.
\newblock Over half of seniors who start oral bisphosphonate therapy are
  exposed for 3 or more years: novel rolling window approach and patterns of
  use.
\newblock {\em Osteoporosis International}, 32(7):1413--1420, 2021.

\bibitem{Hayes2021a}
Kaleen~N. Hayes, Elizabeth~M. Winter, Suzanne~M. Cadarette, and Andrea~M.
  Burden.
\newblock Duration of bisphosphonate drug holidays in osteoporosis patients: A
  narrative review of the evidence and considerations for decision-making.
\newblock {\em Journal of Clinical Medicine}, 10(5), 2021.

\bibitem{hernan2025causal}
M.A. Hernan and J.M. Robins.
\newblock {\em Causal Inference: What If}.
\newblock Chapman \& Hall/CRC Monographs on Statistics \& Applied Probab. CRC
  Press, 2025.

\bibitem{Hernan2000}
Miguel~{\'A}ngel Hern{\'a}n, Babette Brumback, and James~M. Robins.
\newblock Marginal structural models to estimate the causal effect of
  zidovudine on the survival of hiv positive men.
\newblock {\em Epidemiology}, 11(5), 2000.

\bibitem{Hernan2009}
Miguel~A Hernán, Mara McAdams, Nuala McGrath, Emilie Lanoy, and Dominique
  Costagliola.
\newblock Observation plans in longitudinal studies with time-varying
  treatments.
\newblock {\em Statistical Methods in Medical Research}, 18(1):27--52, 2009.

\bibitem{kalbfleisch2011}
John~D Kalbfleisch and Ross~L Prentice.
\newblock {\em The statistical analysis of failure time data}.
\newblock John Wiley \& Sons, 2011.

\bibitem{Rosenbeck2004}
Haiqun Lin, Daniel~O. Scharfstein, and Robert~A. Rosenheck.
\newblock Analysis of longitudinal data with irregular, outcome-dependent
  follow-up.
\newblock {\em Journal of the Royal Statistical Society Series B: Statistical
  Methodology}, 66(3):791--813, 07 2004.

\bibitem{lok2004}
Judith Lok, Richard Gill, Aad Van Der~Vaart, and James Robins.
\newblock Estimating the causal effect of a time-varying treatment on
  time-to-event using structural nested failure time models.
\newblock {\em Statistica Neerlandica}, 58(3):271--295, 2004.

\bibitem{Lok2008}
Judith~J. Lok.
\newblock {Statistical modeling of causal effects in continuous time}.
\newblock {\em The Annals of Statistics}, 36(3):1464 -- 1507, 2008.

\bibitem{Lyu2019}
Houchen Lyu, Sizheng~S Zhao, Kazuki Yoshida, Sara~K Tedeschi, Chang Xu, Sagar~U
  Nigwekar, Benjamin~Z Leder, and Daniel~H Solomon.
\newblock {Comparison of Teriparatide and Denosumab in Patients Switching From
  Long-Term Bisphosphonate Use}.
\newblock {\em The Journal of Clinical Endocrinology \& Metabolism},
  104(11):5611--5620, 07 2019.

\bibitem{naimi2016}
Ashley~I Naimi, Stephen~R Cole, and Edward~H Kennedy.
\newblock {An introduction to g methods}.
\newblock {\em International Journal of Epidemiology}, 46(2):756--762, 12 2016.

\bibitem{Neugebauer2012}
Romain Neugebauer, Bruce Fireman, Jason~A. Roy, Patrick~J. O'Connor, and Joe~V.
  Selby.
\newblock Dynamic marginal structural modeling to evaluate the comparative
  effectiveness of more or less aggressive treatment intensification strategies
  in adults with type 2 diabetes.
\newblock {\em Pharmacoepidemiology and Drug Safety}, 21(S2):99--113, 2012.

\bibitem{oganisian2024}
Arman Oganisian, Kelly~D Getz, Todd~A Alonzo, Richard Aplenc, and Jason~A Roy.
\newblock {Bayesian semiparametric model for sequential treatment decisions
  with informative timing}.
\newblock {\em Biostatistics}, 25(4):947--961, 01 2024.

\bibitem{Lim2016}
Eleanor~M Pullenayegum and Lily~SH Lim.
\newblock Longitudinal data subject to irregular observation: A review of
  methods with a focus on visit processes, assumptions, and study design.
\newblock {\em Statistical Methods in Medical Research}, 25(6):2992--3014,
  2016.

\bibitem{Robins1986}
James Robins.
\newblock A new approach to causal inference in mortality studies with a
  sustained exposure period---application to control of the healthy worker
  survivor effect.
\newblock {\em Mathematical Modelling}, 7(9):1393--1512, 1986.

\bibitem{Robins2000}
James~M. Robins, Miguel~{\'A}ngel Hern{\'a}n, and Babette Brumback.
\newblock Marginal structural models and causal inference in epidemiology.
\newblock {\em Epidemiology}, 11(5), 2000.

\bibitem{Kjetil2011}
Kjetil R{\o}ysland.
\newblock {A martingale approach to continuous-time marginal structural
  models}.
\newblock {\em Bernoulli}, 17(3):895 -- 915, 2011.

\bibitem{Ryalen2018}
Pål~Christie Ryalen, Mats~Julius Stensrud, Sophie Fosså, and Kjetil
  Røysland.
\newblock Causal inference in continuous time: an example on prostate cancer
  therapy.
\newblock {\em Biostatistics}, 21(1):172--185, 08 2018.

\bibitem{sun2023}
Jinghao Sun and Forrest~W Crawford.
\newblock The role of discretization scales in causal inference with
  continuous-time treatment.
\newblock {\em arXiv preprint arXiv:2306.08840}, 2023.

\bibitem{tsiatis2020}
A.A. Tsiatis.
\newblock {\em Dynamic Treatment Regimes: Statistical Methods for Precision
  Medicine}.
\newblock Chapman \& hall/crc monographs on statistics and applied probability.
  CRC Press/Taylor \& Francis Group, 2020.

\bibitem{Young2020}
Jessica~G. Young, Mats~J. Stensrud, Eric~J. Tchetgen~Tchetgen, and Miguel~A.
  Hernán.
\newblock A causal framework for classical statistical estimands in
  failure-time settings with competing events.
\newblock {\em Statistics in Medicine}, 39(8):1199--1236, 2020.

\bibitem{Zhang2011}
Mingyuan Zhang, Marshall~M. Joffe, and Dylan~S. Small.
\newblock {Causal inference for continuous-time processes when covariates are
  observed only at discrete times}.
\newblock {\em The Annals of Statistics}, 39(1):131 -- 173, 2011.

\end{thebibliography}

\newpage
\appendix
\renewcommand{\theequation}{A\thesection.\arabic{equation}}

\section*{Appendix}

\subsection*{A1 - Identification Proofs}
Let  $W_1^{a_1} = \min(W_{A1}^{a_1}, W_{T1}^{a_1})$, $\delta_1^{a_1}=I(W_{A1}^{a_1} < W_{T1}^{a_1}) $, and $W_2^{a_1,a_2}$. So there are $\delta_1(a_1)+1$ \textit{potential} number of treatment courses. The total potential survival time is $T^{a_1,a_2} = W_{1}^{a_1} + \delta_1^{a_1} W_2^{a_1,a_2}$. For some fixed $\tau>0$, the estimand of interest is

 $$ P(T^{a_1, a_2} > \tau) = \int_{\tau}^\infty f_{T^{a_1,a_2}}(t) dt $$

where $f_{T^{a_1,a_2}}(t)$ is the density of the potential survival time. By the total law of probability, this can be expressed as an average over the potential number of treatments
\begin{equation} \label{eq:ai1}
    \begin{split}
        f_{T^{a_1,a_2}}(t) & = \sum_{u=0}^1 f_{T^{a_1,a_2}}(t, \delta_1^{a_1} = u ) \\
                               & = f_{T^{a_1,a_2}}(t, \delta_1^{a_1} = 0 ) + f_{T^{a_1,a_2}}(t, \delta_1^{a_1} = 1 )
    \end{split}
\end{equation}
 The first term captures survival past $\tau$ among those who died before intiated the second course. The second term captures survival past $\tau$ among those who survived through their second treatment course. We will identify each term in \eqref{eq:ai1}, starting from the first term. For the first term note that since $\delta_1^{a_1} = 0$, $T^{a_1,a_2}=W_1^{a_1}$. Below, in the second equality we average over the distribution of confounders measured at the first course, $L_1$. The third equality follows from invoking conditional ignorability of $A_1$ given $L_1$ to condition on $a_1$

\begin{equation*}
    \begin{split}
         f_{T^{a_1,a_2}}(t, \delta_1^{a_1} = 0 )
                        & = \int f_{W_1^{a_1}, \delta_1^{a_1} \mid L_1}(t, \delta_1^{a_1}=0 \mid L_1=l_1) h_{1}(l_1) dl_1 \\
                        & = \int f_{W_1^{a_1}, \delta_1^{a_1} \mid L_1}(t, \delta_1^{a_1}=0 \mid A_1=a_1, L_1=l_1) h_{1}(l_1) dl_1 \\
                        & = \int f_{W_1, \delta_1 \mid L_1}(w_1, \delta_1=0 \mid A_1=a_1, L_1=l_1) h_{1}(l_1) dl_1 \\
    \end{split}
\end{equation*}
 The last equality follows from consistency. Now, in the last line is just $f_{W, \delta_1 \mid L_1}(w_1, \delta_1=0 \mid A_1=a_1, L_1=l_1)$ is just the conditional sub-density of the waiting time from the first course to death, $W_{T1}$ which in the main text we had denoted by $f_{kq}(w_k \mid \bar W_{k-1} = \bar w_{k-1}, \bar A_k= \bar a_k, \bar L_k= \bar l_k)$ with $k=1$ and $q=0$. So, the first term in \eqref{eq:ai1} is

\begin{equation*}
    \begin{split}
         f_{T^{a_1,a_2}}(t, \delta_1^{a_1} = 0 ) & = \int f_{10}(t \mid A_1=a_1, L_1=l_1) h_{1}(l_1) dl_1 \\ 
    \end{split}
\end{equation*}

 Now we identify the second term in a similar fashion starting by averaging over $L_1$ and invoking exchangeability and consistency of $A_1$:

\begin{equation*}
    \begin{split}
        f_{T^{a_1,a_2}}(t, \delta_1^{a_1} = 1 ) 
        & = \int f_{T^{a_1,a_2}}(t, \delta_1^{a_1} = 1 \delta_1^{a_1} = 1 \mid L_1=l_1) h_{1}(l_1) d l_1 \\ 
        & = \int f_{T^{a_1,a_2}}(t, \delta_1^{a_1} = 1 \mid A_1=a_1, L_1=l_1) h_{1}(l_1) d l_1 \\ 
        & = \int \Big[ \int_{0}^t f_{T^{a_1,a_2}}(t \mid W_1 = w_1, A_1=a_1, L_1=l_1,  \delta_1 = 1) \\ 
        & \ \ \ \times f_{W_1, \delta_1 \mid A_1, L_1}(w_1, \delta_1 = 1 \mid A_1=a_1, L_1 = l_1 ) dw_1 \Big] h_{1}(l_1)  dl_1  \\
    \end{split}
\end{equation*}
The last line follows from averaging over the distribution of waiting times which, since $\delta_1=1$, ranges from $0$ to $t$. Now note that $f_{W_1, \delta_1 \mid A_1, L_1}$ is just the observed data sub-density function of the waiting time from the first course to the second course. So, 

\begin{equation*}
    \begin{split}
        f_{T^{a_1,a_2}}(t, \delta_1^{a_1} = 1 ) 
        & = \int \Big[ \int_{0}^t f_{T^{a_1,a_2}}(t \mid W_1 = w_1, A_1=a_1, L_1=l_1,  \delta_1 = 1) \\ 
        & \ \ \ \times f_{11}(w_1\mid A_1=a_1, L_1 = l_1 ) dw_1 \Big] h_{1}(l_1)  dl_1  \\
    \end{split}
\end{equation*}

Averaging over the conditional density of $L_2$, the first factor in brackets above is
\begin{equation*}
    \begin{split}
 f_{T^{a_1,a_2}}(t \mid w_1, a_1, l_1,  \delta_1 = 1) 
 & = \int f_{T^{a_1,a_2}}(t \mid w_1, a_1, \bar l_2,  \delta_1 = 1) h_{2}( l_2 \mid a_1, l_1, w_1 )dl_2 \\
 & = \int f_{T^{a_1,a_2}}(t \mid  w_1, \bar a_2, \bar l_2,  \delta_1 = 1) h_{2}( l_2 \mid a_1, l_1, w_1 )dl_2 \\ 
 & = \int f_{W}(t-w_1 \mid  w_1, \bar a_2, \bar l_2,  \delta_1 = 1) h_{2}( l_2 \mid a_1, l_1, w_1 )dl_2 \\ 
    \end{split}
\end{equation*}
In the first line $h_2$ is the conditional density of $l_2$ among those with $\delta_1=1$. The second line follows from sequential exchangeability. The Third line follows from the fact that $T^{a_1, a_2} = W_1^{a_1} + W_2^{a_1, a_2}$. 

So the second term in \eqref{eq:ai1} is
\begin{equation*}
    \begin{split}
 f_{T^{a_1,a_2}}(t, \delta_1^{a_1} = 1 )  
        & = \int \Big[ \int_{0}^t \Big\{ \int f_{W_2}(t-w_1 \mid  w_1, \bar a_2, \bar l_2,  \delta_1 = 1) h_{2}( l_2 \mid a_1, l_1, w_1 )dl_2\Big\} \\
        & \ \ \ \times f_{11}(w_1\mid a_1, l_1 ) dw_1 \Big] h_{1}(l_1)  dl_1  \\
    \end{split}
\end{equation*}

Now that we have both terms, substituting back into \eqref{eq:ai1} yields the expression in the main text:
\begin{equation*}
    \begin{split}
        f_{T^{a_1,a_2}}(t) & = \int f_{10}(t \mid A_1=a_1, L_1=l_1) h_{1}(l_1) dl_1 \\
        & \ \ \ + \int \Big[ \int_{0}^t \Big\{ \int f_{W_2}(t-w_1 \mid  w_1, \bar a_2, \bar l_2,  \delta_1 = 1) h_{2}( l_2 \mid a_1, l_1, w_1 )dl_2\Big\}f_{11}(w_1\mid a_1, l_1 ) dw_1 \Big] h_{1}(l_1)  dl_1  \\
    \end{split}
\end{equation*}

The survival probability is attained by integrating the density from $\tau$ to infinity, 
\begin{equation*}
    \begin{split}
        P(T^{a_1, a_2} > & \tau)  = \int_{\tau}^\infty\Big( \int f_{10}(t \mid A_1=a_1, L_1=l_1) h_{1}(l_1) dl_1 \\
        & + \int \Big[ \int_{0}^t \Big\{ \int f_{W_2}(t-w_1 \mid  w_1, \bar a_2, \bar l_2,  \delta_1 = 1) h_{2}( l_2 \mid a_1, l_1, w_1 )dl_2\Big\}f_{11}(w_1\mid a_1, l_1 ) dw_1 \Big] h_{1}(l_1)  dl_1 \Big) dt  \\
    \end{split}
\end{equation*}

\subsection*{A2 - Synthetic Data Generation Details}

In this section we describe the data-generating mechanisms used in the simulation analyses and the worked example. For each simmulation scenario, we generate each data sets with $n=2000$ subjects in each data set. For the worked example we generate a small data set of $n=600$ subjects. \\


In the scenario without censoring, the data for a single subject are generated as follows:

\begin{enumerate}
    \item $L_1 \sim Ber(.5)$
    \item $A_1 \mid L_1 \sim Ber\Big( \expit(1 - L_1 ) \Big) $
    \item Generate time to death $ W_{T1} \mid A_1, L_1 \sim Exp\Big( rate = \exp(-3 - A_1 + L_1 ) \Big) $ and time to next treatment $ W_{A1} \mid A_1, L_1 \sim Exp\Big( rate = \exp(-3 + A_1 + L_1 ) \Big) $.
    \item If $W_{T1} < W_{A1}$, then the patient died so set survival time $T=W_{T1}$ and stop. Move on to next patient. Otherwise, the patient survived to treatment 2. Set $W_1 = W_{A1}$ and generate data at second treatment course:
    \item $L_2 \mid L_1, W_1 \sim Ber\Big( \expit( .5 + .15L_1 -.15 I(W_1 > 15) ) \Big)$
    \item $A_2 \mid L_2, W_1 \sim Ber\Big( \expit( 1 - L_2 + 1.5I(W_1>15) ) \Big)$
    \item $W_2 \mid A_2, L_2, W_1 \sim Exp\Big( rate = \exp( -3 + A_2 + L_2 - 2 I(W_1>15) ) \Big)$
    \item The simulated survival time is thus $T=W_1+W_2$.
\end{enumerate}
Notice that subjects may die before reaching the second treatment. \\

In the second scenario, we allow for censoring and so the data for a single subject are generated as follows:

\begin{enumerate}
    \item $L_1 \sim Ber(.5)$
    \item $A_1 \mid L_1 \sim Ber\Big( \expit(1 - L_1 ) \Big) $
    \item Generate time to death $ W_{T1} \mid A_1, L_1 \sim Exp\Big( rate = \exp(-3 - A_1 + L_1 ) \Big) $ and time to next treatment $ W_{A1} \mid A_1, L_1 \sim Exp\Big( rate = \exp(-3 + A_1 + L_1 ) \Big) $.
    \item Generate time to censoring $C_1 \sim Exp\Big( rate  = -4 - A_1 + L_1 \Big)$
    \item If $W_{A1} > \max(W_{T1}, C_1)$, then the subject did not reach the second treatment. Their observed time is $W_1 = \min(C_1, W_{T1})$. The associated indicator is $\delta_1=0$ if they died and $\delta_1=-1$ if they were censored. Stop and move on to next subject. If $W_{A1} < \min(W_{T1}, C_1)$, then the patient started the second treatment. Set the waiting time between treatments to $W_1 = W_{A1}$ and the associated indicator to $\delta_1=1$. Move on to second stage:
    \item $L_2 \mid L_1, W_1 \sim Ber\Big( \expit( .5 + .15L_1 -.15 I(W_1 > 15) ) \Big)$
    \item $A_2 \mid L_2, W_1 \sim Ber\Big( \expit( 1 - L_2 + 1.5I(W_1>15) ) \Big)$
    \item Simulate time from treatment 2 to death: 
    $$ W_{T2} \mid A_2, L_2, W_1 \sim Exp\Big( rate = \exp( -3 + A_2 + L_2 - 2 I(W_1>15) ) \Big) $$
    \item  Simulate time from treatment 2 to censoring:
    $$ C_2 \mid A_2, L_2, W_1 \sim Exp\Big( rate = \exp( -4 + A_2 + L_2 - 2 I(W_1>15) ) \Big) $$
    The observed waiting time is $W_2=\min(W_{T2}, C_2)$. The associated indicator is $\delta_2 = I(W_{T2}<C_2)$.
\end{enumerate}
Repeating the above steps 1,000 times yields a data set for $n=1000$ subjects - who may complete the two-course sequence, die/be censored before second treatment, or be censored before death after the second treatment. Notice that censoring time is allowed to depend on waiting times and confounders. For the worked example, we analyze a data set consisting of $n=600$ patients generated under the second scenario (with censoring).

\subsection*{A3 - Details of Worked Example}
Here we discuss the software details behind the worked example in Section 5 of the main text. \\

To adjust for informative timing we implement the continuous-time approach in Equation \eqref{eq:iptw3} with weights $\hat \omega_i = 1/\hat \pi_i \hat \eta_i$ that are formed using both inverse probability of treatment and censoring weights. We compute the treatment probabilities $\hat \pi_i$ by estimating a logistic regression using the \texttt{glm} function in \texttt{R}. The \texttt{predict} function is used to obtain the fitted probabilities, $\hat \pi_i$, from this estimated logistic regression. The survival probabilities required to compute $\hat\eta_i$ are computed by estimating an exponential proportional hazard (PH) model for the cause-specific hazard of censoring. In the software, censoring time is treated as as the event of interest and the first of either subsequent treatment or death as the ``censoring'' time. For example, with these data we specified an exponential PH model for the waiting time from the first treatment to censoring, $C_1$, as $\lambda(c_1) = \lambda_0(c_1) \exp( \beta_1 A_1 + \beta_2 L_1 )$. The corresponding survival function is given by $S_{C_1}( c_1 \mid A_1, L_1) = \exp( - \Lambda_0(c_1)  \exp( \beta_1 A_1 + \beta_2 L_1 ) )$. Here, $\lambda_0(c_1)$ is the baseline hazard of censoring and $\Lambda_0(c_1) = \int_0^{c_1} \lambda_0(w)dw$ is the cumulative baseline hazard. We use the \texttt{flexsurv} package in \texttt{R} to estimate this PH model with, again for simplicity, a parametric specification of the baseline hazard. \texttt{flexsurv} obtains estimates $\hat \Lambda_0(c_2)$, $\hat \beta_1$, and $\hat \beta_2$ via maximum likelihood estimation. After estimation, we use the \texttt{predict.flexsurvreg} function to plug in these estimates into $S_{C_1}$ to obtain $\hat S_{C_1}( c_1 \mid A_{1i}, L_{1i})$. Alternatively, the function \texttt{coxph} in the \texttt{survival} package can be used to obtain estimates $\hat \beta_1$ and $\hat \beta_2$ via partial likelihood maximization, while a nonparametric Nelson-Aalen estimate of $\Lambda_0(c_1)$ can be obtained via the \texttt{basehaz} function. The function \texttt{predict.coxph} can then be used to compute $ \hat S_{C_1}( c_1 \mid A_{1i}, L_{1i}) = \exp( - \hat \Lambda_0(c_1)  \exp( \hat \beta_1 A_{1i} + \hat \beta_2 L_{1i} ) )$. This provides a more robust alternative.

Using these weights, we estimate $P(T^{1,1} > \tau)$ at points $\tau\in0,1,2,\dots, 20$. Figure \ref{fig:survplot} depicts the estimates where we include waiting time $W_1$ as a covariate in the propensity score and censoring models (in blue) and where we exclude it (in green). The intervals represent 95\% confidence intervals calculated via 300 bootstrap replicates using the \texttt{boot} package.	If parametric approaches are used for estimation of the nuisance parameters, the variance could also be estimated using the sandwich-variance estimator.

As a comparator we also implement the discrete-time adjustment in Equation \eqref{eq:dt_iptw}. This is done as follows. First, the window from time zero to the maximum observed time is partitioned into 1 month intervals up to month 20 - which is the last point on the survival curve we wish to evaluate. We arranged these data into long person-interval format and used a pooled logistic regression, implemented in \texttt{glm}, to estimate the discrete-time hazard of censoring. The cumulative product of these estimates hazards at each interval gives us the estimated probability of remaining uncensored in that interval. We then multiply these with the estimated treatment probabilities given in Section \ref{sc:discrete} which are obtained by separate logistic regressions. The inverse of this product gives us the time-varying weight. Since we are interested in evaluating survival under sequence $(a_1=1, a_2=1)$, we subset to uncensored person-intervals consistent with the dynamic treatment rule, $r$, that assigns $A_k=1$ whenever $V_k=1$. Then fit a logistic regression for outcome $Y_k$ with interval-specific intercepts weighted by the time-varying weights in the \texttt{glm} function. This yields an estimate of the discrete-time hazard $P(Y_j^{\bar d_j} = 1 \mid  Y_{j-1}^{\bar d_{j-1}} = 0 )$, which we then multiply up to desired interval to get the estimated survival up to that interval, as shown in Equation \eqref{eq:dischaz}, using the \texttt{cumprod} function. The results are plotted in black in Figure \ref{fig:survplot}. As can be seen, these results are nearly identical to adjusted estimates using the continuous-time model.

\newpage 

\subsection*{A4 - Graphical Perspective on Sequential Exchangeability}

In Equation \eqref{eq:id3} of the main text discussing the discretized analysis, we required the following sequential conditional exchangeability assumption to hold
\begin{equation*}
    D_j \perp \underline Y_{j+1}^{\bar d_j^r} \mid \bar X_j = \bar x_j,  \bar D_{j-1} = \bar d_{j-1}^r, \bar V_j = \bar v_j, \bar Y_{j} = 0 
\end{equation*}
and described how it is analogous to the required exchangeability assumption in continuous-time. Recall that $V_j$ is an indicator of whether a subject initiated a treatment course at interval $j$. If a course is initiated, then an active treatment, $D_j\in\{0,1\}$, may be given but if it is not initiated then no active treatment is given $D_j=\emptyset$. In this sense, $D_j$ can be viewed as the output of a dynamic treatment rule where $V_j$ is the tailoring variable. \\

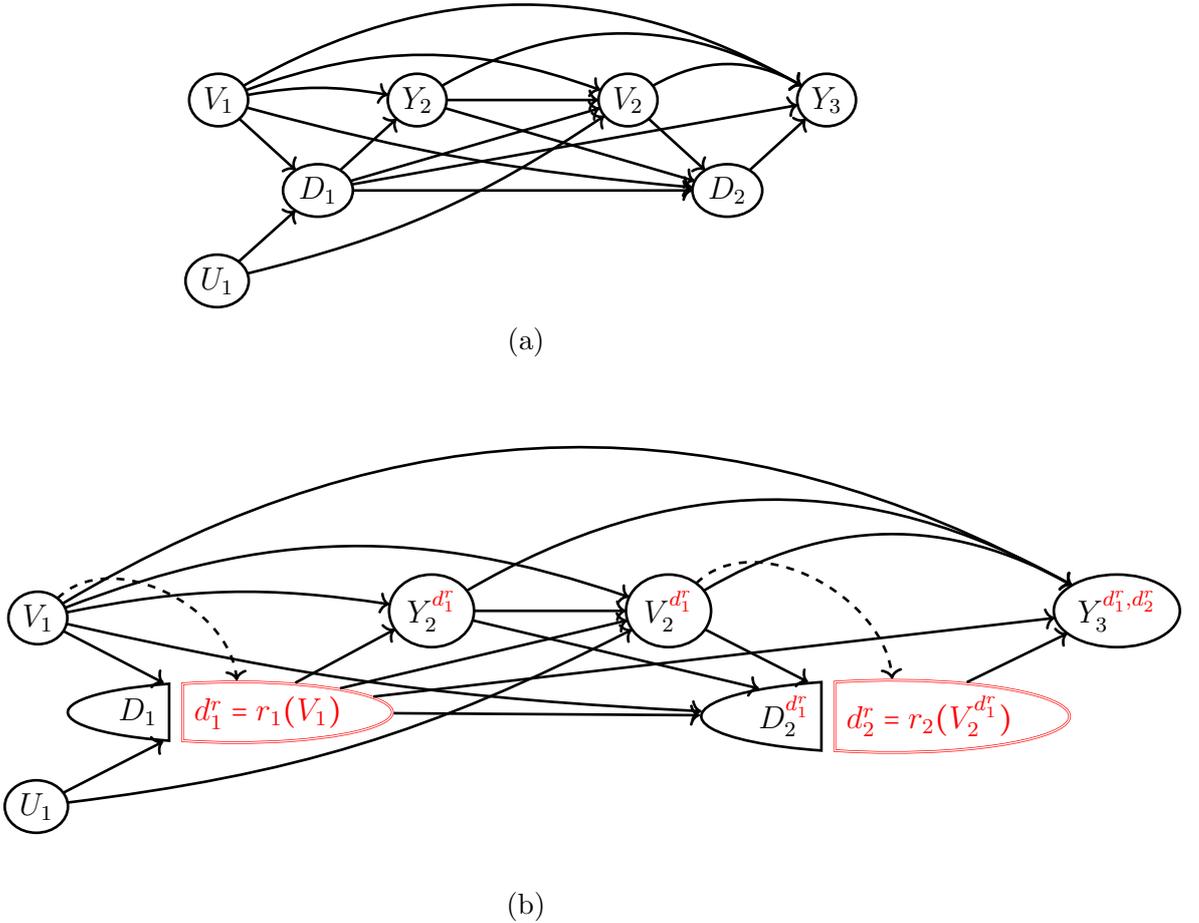
\begin{figure}[h!]
  \centering
  \begin{subfigure}[b]{0.8\textwidth}
    \centering
        \begin{tikzpicture}
\tikzset{line width=1pt, outer sep=0pt, 
         ell/.style={draw,fill=white, inner sep=2pt, line width=1pt}, 
         swig vsplit={gap=5pt, inner line width right=0.5pt, line color right=red}};

\node[name=v1, ell, shape=ellipse]{$V_1$};
\node[name=d1,ell, below right=10mm of v1, shape=ellipse]{$D_1$};
\node[name=u1,ell, below left=10mm of d1, shape=ellipse]{$U_1$};

\node[name=y2, above right = 10mm of d1 , ell, shape=ellipse]{$Y_2$};
\node[name=v2, ell, right = 20mm of y2, shape=ellipse]{$V_2$};
\node[name=d2, ell, below right=10mm of v2, shape=ellipse]{$D_2$};
\node[name=y3, above right = 10mm of d2 , ell, shape=ellipse]{$Y_3$};

\draw[->,line width=1pt]
(u1) edge (d1)
(u1) edge[bend right=10] (v2);

\draw[->,line width=1pt]
(v1) to (d1)
(v1) edge[bend left=10] (y2)
(v1) edge[bend left=20] (v2)
(v1) edge[bend right=5] (d2)
(v1) edge[bend left=30] (y3);

\path[->,line width=1pt]
(d1) edge[bend left=0] (y2)
(d1) edge[bend right=0] (v2)
(d1) edge[bend left=0] (d2)
(d1) edge[bend right=0] (y3);

\path[->,line width=1pt]
(y2) edge[bend left=0] (v2)
(y2) edge[bend left=0] (d2)
(y2) edge[bend left=30] (y3);

\draw[->,line width=1pt]
(v2) to (d2)
(v2) edge[bend left=30] (y3);

\path[->, line width = 1pt] (d2) edge (y3);

\end{tikzpicture}
    \caption{}
  \end{subfigure}
  
  \vspace{1em} 
  
  \begin{subfigure}[b]{0.8\textwidth}
    \centering
        \begin{tikzpicture}
\tikzset{line width=1pt, outer sep=0pt, 
         ell/.style={draw,fill=white, inner sep=2pt, line width=1pt}, 
         swig vsplit={gap=5pt, inner line width right=0.5pt, line color right=red}};

\node[name=v1, ell, shape=ellipse]{$V_1$};
\node[name=d1, below right=10mm of v1, shape=swig vsplit]{
    \nodepart{left}{$D_1$}
    \nodepart[text=red]{right}{$d_1^r=r_1(V_1)$} };
\node[name=u1,ell, below left=10mm of d1, shape=ellipse]{$U_1$};
\node[name=y2, above right = 10mm of d1 , ell, shape=ellipse]{$Y_2^{\textcolor{red}{d_1^r}}$};
\node[name=v2, ell, right = 20mm of y2, shape=ellipse]{$V_2^{\textcolor{red}{d_1^r}}$};
\node[name=d2, below right=10mm of v2, shape=swig vsplit]{
    \nodepart{left}{$D_2^{\textcolor{red}{d_1^r}}$}
    \nodepart[text=red]{right}{$d_2^r=r_2(V_2^{d_1^r})$} };
\node[name=y3, above right = 10mm of d2 , ell, shape=ellipse]{$Y_3^{ \textcolor{red}{d_1^r, d_2^r}}$};

\draw[dashed, ->, line width=1pt] (v1) to[in=north] (d1);
\draw[->,line width=1pt]
(v1) to (d1)
(v1) edge[bend left=10] (y2)
(v1) edge[bend left=20] (v2)
(v1) edge[bend right=5] (d2)
(v1) edge[bend left=30] (y3)
;

\draw[->,line width=1pt]
(u1) edge (d1)
(u1) edge[bend right=10] (v2);

\path[->,line width=1pt]
(d1) edge[bend left=0] (y2)
(d1) edge[bend right=0] (v2)
(d1) edge[bend left=0] (d2)
(d1) edge[bend right=0] (y3)
;

\path[->,line width=1pt]
(y2) edge[bend left=0] (v2)
(y2) edge[bend left=0] (d2)
(y2) edge[bend left=30] (y3)
;

\draw[dashed, ->, line width=1pt] (v2) to[in=north] (d2);
\draw[->,line width=1pt]
(v2) to (d2)
(v2) edge[bend left=30] (y3)
;

\path[->, line width = 1pt] (d2) edge (y3);
\end{tikzpicture}
    \caption{}
  \end{subfigure}  
  \caption{(a) DAG of discrete-time process described in Section \ref{sc:discrete}, where $V_1$ is an indicator of a treatment course initiated at course 1, $D_1$ indicating the treatment decision made, and $Y_j$ denoting survival status. We abbreviation the DAG for $j=1, 2$ rather than $j=1,2,\dots, J$ and omit measured features $X_j$ for compactness. Here, $U_1$ denotes unmeasured common causes of treatment decision and whether someone subsequently initiates the next course. (b) A Single World Intervention Graph (SWIG) template for rule $r = \{r_1, r_2, \dots, r_J\}$ corresponding to the DAG in the top panel. Rule $r_j$ maps to one of two active treatments if a course is initiated ($V_j=1$) and maps to an inactive treatment $d_1^r=\emptyset$ if a subject is not ready to initiate treatment. Dashed arrows indicate additional arrows from the tailoring variable $V_j$ to the decision node.}
  \label{fig:swig}
\end{figure}

With static treatment rules, sequential exchangeability asserts that $(\bar X_j ,  \bar D_{j-1}, \bar V_j, \bar Y_{j})$ d-separates $D_j$ and $\underline Y_{j+1}^{\bar d_j^r}$ in a single-world intervention graph (SWIG). In the case of dynamic treatment rules, violations can occur if there are 1) unblocked backdoor paths from $D_j$ to future values of the tailoring variable as well and 2) unblocked backdoor paths from $D_j$ to future $V_j$. These points are discussed at length in Chapter 19 of \cite{hernan2025causal}. Interested readers should see especially Section 19.5. We provide a brief summary of these points for this particular application with informative timing. \\

To illustrate this, Figure 3 shows a possible DAG and the corresponding SWIG template for DTR $r=\{r_1, r_2,\dots,r_J\}$. That is, the SWIG represents the joint distribution of these random variables in a world in which we intervene to set treatment decisions based on a rule $r$ with tailoring variable process ${V_j}$. We have omitted observed covariate process, $\{X_j\}$ and only depicted the graphs for $j\in\{1,2\}$ for compactness and to focus attention on ${V_j}$. We assume everyone enters the study alive so that $Y_1:=1$ for all subjects. For $j=1,2$ exchangeability asserts that

\begin{equation*}
    \begin{split}
        D_1 \perp ( Y_3^{d_1^r, d_2^r}, Y_2^{d_1^r})   \mid  V_1=v_1, Y_1=0 \\
        D_2 \perp Y_3^{d_1^r, d_2^r}   \mid  D_1=d_1^r, \bar V_2= \bar v_2, Y_2=0 \\
    \end{split}
\end{equation*}

Since $V_j$ is a tailoring variable, new dashed paths, absent on the DAG Figure \ref{fig:swig}(a), now appear on the SWIG in Figure \ref{fig:swig}(b). Note in the second line above we omit the superscripts on $D_2$ and $V_2$ which are present on the SWIG by consistency, since $d_1^r$ is conditioned on. In the presence of unmeasured common causes of treatment decision and course initiation, such as $U_1$, $D_1$ and $Y_2^{\bar d_2^r}$ are not d-separated due to open backdoor path $D_1 \leftarrow U_1 \rightarrow V_2^{d_1^r} \dashedrightarrow  r_2(V_2^{d_1^r}) \rightarrow Y_3^{d_1^r, d_2^r}$. Thus, sequential exchangeability is violated since $D_1 \not \perp ( Y_3^{d_1^r, d_2^r}, Y_2^{d_1^r})  \mid  V_1=v_1, Y_1=0$. This is true even if there is no direct effect of treatment course initiation on outcomes (arrows from $V_j \rightarrow Y_{j+1}$) on the DAG.

\end{document}